\documentclass[12pt]{iopart}

\usepackage{iopams}
\usepackage{graphicx}
\usepackage{epsfig}
\usepackage{enumerate}
\usepackage{cite}

\newcommand{\CD}{{\cal D}}
\newcommand{\CQ}{{Q}}

\newcommand{\CW}{{\cal W}}
\newcommand{\CR}{{R}}

\newcommand{\Omalpha}{{\Omega_{\alpha}}}
\newcommand{\Ombeta}{{\Omega_{\beta}}}
\newcommand{\Omgamma}{{\Omega_{\gamma}}}

\newcommand{\average}[1]{\left\langle #1 \right\rangle_\CD}

\newcommand{\initial}[1]{{#1_{\it i}}}

\begin{document}

\title[Chaplygin Gas and Inhomogeneous universe]{Chaplygin gas and effective description of inhomogeneous universe models in general relativity}

\medskip

\author{Xavier Roy and Thomas Buchert}

\address{Universit\'e Lyon 1, Centre de Recherche Astrophysique de Lyon, CNRS UMR 5574 \\
9 avenue Charles Andr\'e, F--69230 Saint--Genis--Laval, France \\
\medskip
Emails: roy@obs.univ-lyon1.fr and buchert@obs.univ-lyon1.fr}

\begin{abstract}
In the framework of spatially averaged inhomogeneous cosmologies in classical general relativity, effective Einstein equations govern the dynamics of averaged scalar variables in a scale--dependent way. A particular cosmology may be characterized by a cosmic equation of state, closing the hierarchy of effective equations. In this context a natural candidate is provided by the Chaplygin gas, standing for a unified description of dark energy and dark matter. In this paper, we suppose that the inhomogeneous properties of matter and geometry obey the Chaplygin equation of state. The most extreme interpretation assumes that both dark energy and dark matter are not included as additional sources, but are both manifestations of spatial geometrical properties. This feature is an important conceptual difference in comparison with the standard approach of a Friedmann--Lema\^itre--Robertson--Walker universe filled with dust and another fundamental field characterized by the Chaplygin equation of state. We finally discuss the consequences of the resulting scenario for effective cosmological parameters in order to establish the framework of a future confrontation with observations, and we note that the standard Chaplygin gas may not be ruled out by them.
\end{abstract}


\pacs{04.20.-q, 04.40.-b, 95.35.+d, 95.36.+x, 98.80.-Es, 98.80.-Jk}


\section{Introduction}

Does an inhomogeneous universe evolve on average like a homogeneous solution in the framework of general relativity? This question is not new \cite{ellis:average} and naturally emerges in view of the nonlinearity of the theory and, in particular, from the generally non--commuting operations of averaging and time evolution \cite{ellisbuchert}. The main difficulty to answer it resides in the notion of averaging and in its construction (see, e.g., \cite{buchert:review}, section 2.2 of \cite{behrend},~and references therein).

Our universe is supposed to verify the strong cosmological principle which demands homogeneity and isotropy at all scales. This standard approach, known as Friedmann--Lema\^itre--Robertson--Walker (FLRW) cosmology,~is widely used in order to describe the dynamics of our universe and the formation of its constituents. It however leaves in suspense an explanation about the origin of dark energy and dark matter,~which respectively represent in this model about $3/4$ and $1/4$ of the total content of the universe. This last point might actually reveal a symptom of a deeper problem linked to this approach. Indeed,~in FLRW cosmology one determines background quantities regardless of the scale and makes them evolve according to a homogeneous--isotropic solution of Einstein equations. Our first query could be reformulated in order to note the central aspect of this issue: Are the background quantities well defined within standard cosmology,~i.e.\ as a suitable average over the inhomogeneities? Is their {\it evolution} well approximated in this framework,~i.e.\ is the time dependence of the homogeneous--isotropic {\it averaged state} well approximated by a homogeneous--isotropic {\it solution}?

We shall adopt an approach that averages, in a domain--dependent way,~the scalar parts of Einstein's equations with respect to synchronous free--falling observers in a dust model \cite{buchert:dust,buchert:fluid},~a realization of the averaging problem that does not answer the above questions in the affirmative. The average evolution of an inhomogeneous universe differs from the evolution of a homogeneous one;~in other words,~even if we are entitled to describe structure formation in terms of perturbations of a background,~this latter is generally not a member of the homogeneous solutions (see also \cite{kolb:voids,kolb:backgrounds}). This difference of evolution is driven by the non--trivial geometrical structure of an inhomogeneous space,~featuring deviations that are known as ``backreaction''. These backreaction effects can act on average,~at least qualitatively,~as the dark components.

The set of equations obtained within this approach should be closed to derive the evolution of all the involved quantities,~namely the effective scale factor,~the averaged scalar curvature deviation and the kinematical backreaction variable. In recent papers, attention was turned to a closure under the assumption of global constraints such as a globally stationary universe \cite{buchert:darkenergy,buchert:static},~or by exploring the solution space with exact scaling laws for the backreaction and the averaged scalar curvature \cite{morphon,morphon:obs},~or by symmetry requirements such as spherical symmetry (e.g.~\cite{bolejkoandersson,celerieretal,enqvist,kolb:lightcone,paranjape,sussman}, and references therein). In this work we want to choose the closure relation by focusing on the particularity of the model to unify the dark components through backreaction.
According to this point of view,~the Chaplygin gas (CG) seems to be an interesting lead since it unifies dark matter and dark energy in only one fluid, obeying an exotic equation of state \cite{gor:demodel,gor:demodel-2,kam:alter}. This unification is made through the evolution of this particular fluid and it can be extended to a unification where both dark components are {\it simultaneously} modelled thanks to the scale dependence of our approach\footnote{A `simultaneous' unification has also been proposed in the context of an inhomogeneous, but fundamental CG in \cite{bilic:inhom_cg}.}. These points motivate us to build a model in which the generically existing coupling between the backreaction and the averaged scalar curvature deviation,~which encodes the particular geometrical structure evolution of an inhomogeneous universe,~is furnished by a scale--dependent CG equation of state\footnote{A study of inhomogeneous spherically symmetric spacetimes, presenting nonlinear perturbations constructed from the fluctuations of local variables with respect to background quantities called quasi--local scalars, has been given in \cite{sussman-chap}. As an example the CG is employed in \cite{sussman-chap} and relates the quasi--local variables' pressure and density. The reader may find connections between quasi--local variables and our averaging procedure in \cite{sussman-ql}.}.

\medskip

In section \ref{sec-inhomcosm} we introduce the equations that govern the average evolution of an inhomogenous universe model and briefly discuss them. We present in section~\ref{sec-CG} the CG and some of its properties. In section \ref{sec-bCG} we suppose that the backreaction and the averaged scalar curvature deviation are coupled according to the CG equation. The CG then describes a particular geometrical structure of the inhomogeneous universe and does not correspond to any fundamental field. Finally,~in section~\ref{sec-cosparm},~we reformulate the results obtained in terms of effective cosmological parameters;~we compare this model to the Friedmannian framework and we study in particular the acceleration of a spatial domain.


\section{Effective description of inhomogeneous universe models} \label{sec-inhomcosm}

Restricting attention to a universe filled with irrotational dust,~i.e.\ irrotational pressureless matter,~we spatially average the scalar parts of Einstein equations (the Ha\-mil\-to\-nian constraint, Raychaudhuri's equation and the continuity equation) with respect to a collection of comoving (generalized fundamental) observers over a compact,~restmass preserving spatial domain $\CD$,~and obtain the following set of equations (\cite{buchert:dust,buchert:fluid,buchert:review}, \cite{rasanen:acceleration} for details):

\vspace{-5mm}
\begin{eqnarray}
	\left(\frac{{\dot a}_\CD}{a_\CD}\right)^2 - \, \frac{8 \pi G}{3} \average{\varrho} \, = \, -\frac{\average{\CR} + {\CQ}_\CD}{6} \, , \label{eff-fried-1} \\
	\frac{{\ddot a}_\CD}{a_\CD} \;\; + \;\; \frac{4 \pi G}{3} \average{\varrho} \, = \, \frac{{\CQ}_\CD}{3} \, , \label{eff-fried-2} \\
	\langle{\varrho}\rangle\dot{}_\CD \;\; + \;\; 3 \, \frac{{\dot a}_\CD}{a_\CD} \average{\varrho} \, = \, 0 \, , \label{eff-fried-3} \\
	\frac{1}{a_\CD^6} \,  \left(\,{\CQ}_\CD \, a_\CD^6 \,\right)\dot{} \;+\; \frac{1}{a_\CD^{2}} \, \left(\average{\CR} \, a_\CD^2 \, \right)\dot{} \, = \, 0 \, , \label{eff-fried-4}
\end{eqnarray}

\noindent
where $a_\CD$ is the effective volume scale factor

\begin{equation}
\label{scale-factor}
   a_\CD (t) := \left( \frac{V_{\CD}(t)}{V_{\initial\CD}} \right)^{1/3} \, ,
\end{equation}

\noindent
with $V_{\initial\CD}$ the initial volume of the domain and $V_{\CD}(t)$ its volume at a proper time $t$, $\average{\varrho} = M  \, a_\CD^{-3} / V_{\initial\CD}$ the density of irrotational dust averaged over $\CD$, $\average{\CR}$ the spatial scalar curvature averaged over $\CD$ and ${\CQ}_\CD$ the kinematical backreaction

\begin{equation}
\label{backreactionterm}
  {\CQ}_\CD (t) := \frac{2}{3}\average{\left(\theta - \average{\theta}\right)^2 } - 2\average{\sigma^2} \, ,
\end{equation}

\noindent
with $\theta$ the rate of expansion and $\sigma:= \sqrt{\frac{1}{2} \sigma^{ij}\sigma_{ij}}$ the rate of shear with the shear tensor components $\sigma_{ij}$.

Equations (\ref{eff-fried-1}) and (\ref{eff-fried-2}) govern the kinematics of the effective scale factor and equations (\ref{eff-fried-3}) and (\ref{eff-fried-4}) express the conservation law for the dust matter and the backreaction terms, respectively. It is important to point out that $\average{\CR}$ might evolve differently from its Friedmannian counterpart (as we shall discuss in subsection \ref{subs-limitW}). Upon introducing the averaged scalar curvature deviation $\CW_\CD := \average{\CR} - 6 k_{\CD_{\it i}} a^{-2}_\CD$,~we may rewrite equation (\ref{eff-fried-1}) in the form

\begin{equation}
	\left(\frac{{\dot a}_\CD}{a_\CD}\right)^2 + \, \frac{k_{\CD_{\it i}}}{a^2_\CD} \, - \, \frac{8 \pi G}{3} \average{\varrho} \, = \, -\frac{\CW_\CD + {\CQ}_\CD}{6} \, . \label{eff-fried-5}
\end{equation}

\noindent
Looking now at equations (\ref{eff-fried-2}) and (\ref{eff-fried-5}) one should note that both the backreaction and the averaged curvature, through $\CW_\CD$, induce a change in the averaged dynamics of the domain in comparison with the Friedmannian framework. Equation (\ref{eff-fried-2}) states that a positive backreaction contributes to accelerate the expansion of the domain and then plays against gravity: ${\CQ}_\CD > 0 \,$ effectively mimics a dark energy behaviour over $\CD$. The domain will actually undergo an acceleration of its expansion only if the `intensity' of dark energy is sufficient,~which is the case when ${\CQ}_\CD > 4 \pi G \langle{\varrho}\rangle_\CD$. A negative backreaction contributes to decelerate the domain expansion and therefore adds to gravity: ${\CQ}_\CD < 0 \,$ effectively mimics a dark matter behaviour over $\CD$. For the averaged model we may suppose that backreaction acts as dark matter on small scales (e.g.\ galaxy cluster and void scales) and as dark energy on the largest scales (CMB and high--redshift supernovae). In the present work, this differentiation with respect to the spatial scale will however not be made explicit. An explicit multi--scale dynamics can be formulated to refine such a description \cite{multiscale}.

\bigskip

We shall assimilate here the properties of the spatial geometrical structure to a {\it domain--dependent} CG. To this aim we first describe the backreaction variables in terms of an effective perfect fluid whose energy density and pressure read\footnote{In this work we prefer to consider the deviation term $\CW_\CD$ to describe the fluid, instead of the full averaged scalar curvature $\average{\CR}$. First, $(\CQ_\CD,\CW_\CD)$ incorporate the deviation from a general Friedmannian model, being equivalent to the pair $(\CQ_\CD,\average{\CR})$ only in a zero--curvature Friedmannian model. Second, the kinematical backreaction and the curvature deviation both vanish on the background and are gauge--invariants (as shown to second--order in perturbation theory \cite{gaugeinv}).}

\begin{equation}
	\varrho^{\CD}_{b} = - \frac{1}{16 \pi G} \, ( \, {\CQ}_\CD + \CW_\CD \, ) \, , \qquad {p}^{\CD}_{b} = - \frac{1}{16 \pi G} \, ( \, {\CQ}_\CD - \frac{\CW_\CD}{3} \, ) \, . \label{back-fluid}
\end{equation}

\noindent
We stress here that, since it is an effective description, this fluid does not have to satisfy any energy conditions (as discussed in \cite{morphon}). We therefore reformulate equations (\ref{eff-fried-1}) and (\ref{eff-fried-2}) casting them into Friedmannian form

\vspace{-5mm}
\begin{eqnarray}
	\left(\frac{{\dot a}_\CD}{a_\CD}\right)^2 \, + \, \frac{k_{\CD_{\it i}}}{a^2_\CD} \, - \, \frac{8 \pi G}{3} \, \left( \average{\varrho} + \varrho^{\CD}_{b} \right) \, = \, 0 \, , \label{eff-fried-6} \\
	\frac{{\ddot a}_\CD}{a_\CD} + \, \frac{4 \pi G}{3} \, \left( \average{\varrho} + \varrho^{\CD}_{b} + 3 {p}^{\CD}_{b} \right) \, = \, 0 \, . \label{eff-fried-7}
\end{eqnarray}

\noindent
Using the last two equations together with equation (\ref{eff-fried-3}) one obtains the conservation law for the backreaction fluid

\begin{equation}
	\label{backreactioncontinuity}
	{\dot\varrho}^{\CD}_{b} + 3 \, \frac{{\dot a}_\CD}{a_\CD} \, \left(\varrho^{\CD}_{b} + {p}^{\CD}_{b} \right) = 0 \, ,
\end{equation}

\noindent
which, if written out, reflects the generic coupling between the curvature deviation and the kinematical backreaction---a simple reformulation of equation (\ref{eff-fried-4}):

\begin{equation}
	\label{integrability}
	\frac{1}{a_\CD^6} \,  \left(\,{\CQ}_\CD \, a_\CD^6 \,\right)\dot{} \;+\; \frac{1}{a_\CD^{2}} \, \left(\CW_\CD \, a_\CD^2 \, \right)\dot{} \,= 0 \, .
\end{equation}


\section{The Chaplygin gas} \label{sec-CG}

The CG is a perfect fluid obeying the state equation

\begin{equation}
	{p}_{ch} = - \frac{A}{\varrho_{ch}} \, , \label{chapyglingas}
\end{equation}

\noindent
where $\varrho_{ch} > 0$ and ${p}_{ch}$ are respectively the energy density and the pressure of the fluid in a comoving frame and $A$ is a positive constant. It was first introduced as a cosmological fluid unifying dark matter and dark energy by Kamenshchik {\it et al} \cite{kam:alter} and has been since widely studied in this context (see, e.g., \cite{avel,bean,bilic,bilic:tach,cart,chim-2,chim,dev,fabris,fabris-2,gor:demodel,gor:demodel-2}). The equation of state (\ref{chapyglingas}) has also raised interest in particle physics thanks to its connection with string theory \cite{bord} and its supersymmetric extension \cite{jack}. The generalization of the CG \cite{bento,kam:alter},

\begin{equation} \label{genCG}
	{p}_{ch} = - \frac{A}{\varrho_{ch}^{\alpha}} \, ,
\end{equation}

\noindent
with $\alpha$ a free positive parameter,~is commonly used in cosmological models;~however in the present work we shall consider for simplicity the case $\alpha = 1$,~i.e.\ the standard CG\footnote{A generalization of our ideas, using equation (\ref{genCG}), is straighforward. We emphasize that, even if the standard CG does not seem to well fit with observations in a FLRW model, this must not be the case in our approach, since observational data have to be reinterpreted before the need for such a generalization is justified (cf section \ref{sec:concl}).}. Assuming that the gas verifies the energy conservation law over a spatial domain $\CD$,

\begin{equation}
\label{cg-cont}
	{\dot\varrho}^{\CD}_{ch} + 3 \, \frac{{\dot a}_\CD}{a_\CD} \, \left(\varrho^{\CD}_{ch} + {p}^{\CD}_{ch} \right) = 0 \, ,
\end{equation}

\noindent
we obtain,~making use of relation (\ref{chapyglingas}),~the expressions

\begin{equation} \label{chap_evol}
	\varrho^{\CD}_{ch} = \sqrt{A_{\CD} + \frac{B_{\CD}}{a^6_\CD}} \, , \qquad {p}^{\CD}_{ch} = - \frac{A_{\CD}}{\sqrt{A_{\CD} + B_{\CD}/a^6_\CD}} \, , \label{cgevol}
\end{equation}

\noindent
where $B_{\CD} = \varrho^{\CD^{\, \scriptstyle 2}}_{{ch}_{\, \scriptstyle i}} - A_{\CD}$ determines the initial conditions of the CG and both $A_{\CD}$ and $B_{\CD}$ depend on the domain. Note that equations (\ref{chap_evol}) describe the evolution of a homogeneous CG which is the one of interest in our work since the backreaction terms are, due to the averaging procedure,~homogeneous over a spatial domain. However, since we average over inhomogeneities, there certainly exist interesting links to the inhomogeneous CG (for an investigation of the latter the reader is referred to \cite{bilic:inhom_cg}).

We now recall briefly of the different aspects of the CG according to the sign of $B_{\CD}$.


\subsection{Choosing a positive integration constant} \label{posB}

We follow \cite{kam:alter} to present the behaviour of the CG for a positive $B_{\CD}$. Expressions~(\ref{cgevol}) become for small values of the scale factor,~$a^6_\CD \ll B_{\CD}/A_{\CD}$,

\begin{equation}
	\varrho^{\CD}_{ch} \sim \frac{\sqrt{B_{\CD}}}{a^3_\CD} \, , \qquad {p}^{\CD}_{ch} \sim 0 \, , \label{cgdm}
\end{equation}

\noindent
which indicates that the CG can behave as a dark matter component. For large values of the scale factor, $a^6_\CD \gg B_\CD/A_\CD$, it follows that

\begin{equation}
	\varrho^{\CD}_{ch} \sim \sqrt{A_\CD} \, , \qquad {p}^{\CD}_{ch} \sim - \sqrt{A_\CD}  \, , \label{cgde}
\end{equation}

\noindent
which reflects the dark energy--like behaviour of the CG in its last stage. Finally, one may also develop (\ref{cgevol}) for large values of $a_\CD$ to obtain

\begin{equation}
	\varrho^{\CD}_{ch} \sim \sqrt{A_\CD} + \frac{B_\CD}{2\sqrt{A_\CD}} a_\CD^{-6} \, , \qquad {p}^{\CD}_{ch} \sim - \sqrt{A_\CD} + \frac{B_\CD}{2\sqrt{A_\CD}} a_\CD^{-6} \, .
\end{equation}

\noindent
Between the phases (\ref{cgdm}) and (\ref{cgde}) the CG can be seen as a mixture of a cosmological constant and a stiff fluid whose pressure and energy density are equal.

To resume,~for a positive $B_\CD$, the CG acts first as dark matter and then as dark energy whose state equation evolves towards the one of a cosmological constant.


\subsection{Choosing a negative integration constant} \label{negB}

As already noticed in \cite{set} the CG presents another interesting feature for a negative $B_{\CD}$ since its density increases with the scale factor. In this situation it plays the role of phantom dark energy, and it evolves at late times towards a cosmological constant. For the pressure and energy density to be well defined one needs

\begin{equation}
	A_{\CD} + \frac{B_{\CD}}{a^6_\CD} \, > \, 0 \;\; \Leftrightarrow \;\; a^6_\CD \, > \, - \frac{B_{\CD}}{A_{\CD}} \;\, .
\end{equation}

\noindent
It therefore exists a minimal value for the scale factor, $a^{\mathrm{min}}_\CD = (- B_{\CD} / A_{\CD} )^{1/6}$,~implying that this case describes a bouncing universe model at early times.


\section{Evolution of the kinematical backreaction and the curvature deviation} \label{sec-bCG}


\subsection{Backreaction fluid as a Chaplygin gas}

As we have outlined in section~\ref{sec-inhomcosm}, the backreaction fluid inherits a {\it simultaneous} unification of the dark components thanks to its scale dependence. We want to build a model in which this fluid is assimilated to the Chaplygin gas in order to physically shape the behaviour of the backreaction and the curvature deviation on a given spatial scale, while also allowing for a metamorphosis of the `dark character' through evolution. In this model, the Chaplygin gas emerges from the inhomogeneous structure of the universe, and is not related to any fundamental field. We then consider that the backreaction fluid responds to the scale--dependent state equation (\ref{chapyglingas}) over a spatial domain $\CD$. Using definitions (\ref{back-fluid}),~we rewrite expression (\ref{chapyglingas}) in the form

\begin{equation} \label{Chapconst}
	({\CW}_\CD + {\CQ}_\CD)(\frac{{\CW}_\CD}{3} - \, {\CQ}_\CD) = (16 \pi G )^2 A_\CD \, . \label{bf-cst}
\end{equation}

\noindent
Since no energy condition has to be verified by the backreaction fluid, we may also consider the Chaplygin equation of state (\ref{chapyglingas}) with a negative energy density (and hence a negative $A_\CD$ if one wants to preserve the negativity of the pressure). However,~we shall restrict ourselves in this paper to the case where the backreaction fluid satisfies (\ref{chapyglingas}) under the usual conditions, i.e.\ with a positive energy density and a positive $A_\CD$\footnote{We do not impose here any other energy conditions on the backreaction fluid.}. We thus have to respect, in view of expression (\ref{Chapconst}), the following constraints, which we call the {\it Chaplygin fluid constraints}:

\begin{equation} \label{ChapconstI}
	{\CW}_\CD + {\CQ}_\CD \, < \, 0 \; ; \qquad \frac{{\CW}_\CD}{3} - \, {\CQ}_\CD \, < \, 0 \, ,
\end{equation}

\noindent
or, equivalently,

\begin{equation}
	{\CW}_\CD \, < \, 0 \, , \qquad \frac{{\CW}_\CD}{3} \, < \, {\CQ}_\CD \, < \, - {\CW}_\CD \, .
\end{equation}

\begin{figure}[ht]
	\begin{center}
	\vspace{1cm}
	\includegraphics[scale=1]{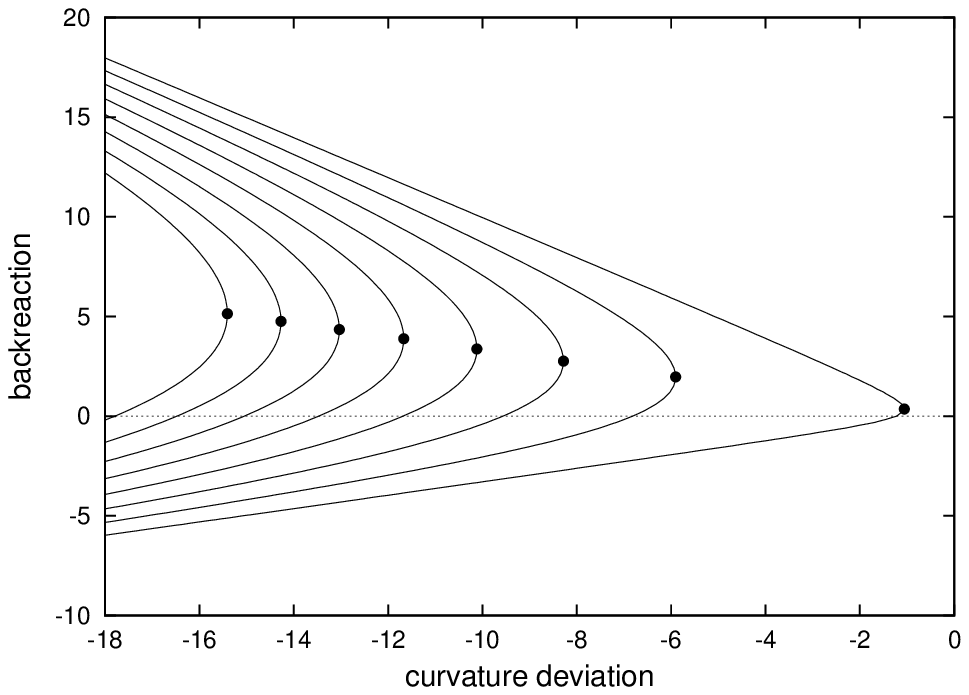}
	\end{center}
	\caption{Each curve draws relation (\ref{bf-cst}) for a different $A_\CD$ (its value arithmetically increases from the curves on the right to the ones on the left) under the Chaplygin fluid constraints. For a given $A_\CD$,~whatever their evolution could be,~${\CQ}_\CD$ and ${\CW}_\CD$ run on the corresponding iso--$A_\CD$ curve. The dots represent the maximal value of the curvature deviation for each $A_\CD$. A negative ${\CQ}_\CD$ mimics dark matter and a positive ${\CQ}_\CD$ dark energy.}
	\label{fig:iso-A}
\end{figure}

\noindent
In this situation, the curvature deviation of any domain is negative at any time whatever its dynamics could be. We also note that ${\CQ}_\CD$ and ${\CW}_\CD$ evolve in such a way that relation (\ref{bf-cst}) is always satisfied (see figure \ref{fig:iso-A}). Rewriting equation (\ref{bf-cst}) as a function ${\CW}_\CD({\CQ}_\CD,A_\CD)$ under the Chaplygin fluid constraints,~we derive,~for a given $A_\CD$,~the maximal value of the deviation term $\CW_M = -24 \pi G \sqrt{A_\CD}$ obtained for $Q_{\CW_M} = 8 \pi G \sqrt{A_\CD}$. We shall see in subsections \ref{subs-limitW} and \ref{subs-limitQ} that these values form an attractor for the dynamics of the system.

\subsection{Exact evolution equations} \label{subs-evoleq}

Inserting expressions (\ref{back-fluid}) into the evolution laws (\ref{cgevol}) results in

\vspace{-5mm}
\begin{eqnarray}
	{\CW}_\CD & = & - \frac{24 \pi G}{\sqrt{A_\CD + B_\CD/a_\CD^6}} \; ( A_\CD + \frac{B_\CD}{2a_\CD^6} \, ) \, , \label{bfevol-r} \\
	{\CQ}_\CD & = & \frac{8 \pi G}{\sqrt{A_\CD + B_\CD/a_\CD^6}} \; ( A_\CD - \frac{B_\CD}{2a_\CD^6} \, ) \, . \label{bfevol-q}
\end{eqnarray}

\noindent
We may express the constants $A_\CD$ and $B_\CD$ in terms of the initial values of the backreaction $\initial{\CQ}$ and the curvature deviation $\initial{\CW}$ over $\CD$,

\vspace{-5mm}
\begin{eqnarray}
	A_\CD & = & \frac{1}{256 \pi^2 G^2} \, (\initial{\CW} + \initial{\CQ} ) \, ( \frac{\initial{\CW}}{3} - \initial{\CQ} ) \, , \label{Ainit} \\
	B_\CD & = & \frac{1}{128 \pi^2 G^2} \, (\initial{\CW} + \initial{\CQ} ) \, ( \frac{\initial{\CW}}{3} + \initial{\CQ} ) \, , \label{Binit}
\end{eqnarray}

\noindent
and determine the evolution of ${\CW}_\CD$ and ${\CQ}_\CD$ in the form

\vspace{-5mm}
\begin{eqnarray}
	{\CW}_\CD & = & -\frac{3}{2} \; \frac{\alpha \beta + \alpha \gamma \, a_\CD^{-6} }{\left( \alpha \beta + 2 \, \alpha \gamma \, a_\CD^{-6} \, \right)^{1/2}} \, , \label{evol-r} \\
	{\CQ}_\CD & = & \; \frac{1}{2} \; \frac{\alpha \beta - \alpha \gamma \, a_\CD^{-6} }{\left( \alpha \beta + 2 \, \alpha \gamma \, a_\CD^{-6} \, \right)^{1/2}} \, , \label{evol-q}
\end{eqnarray}

\noindent
where the new terms are defined as

\begin{equation}
	\fl \alpha := \initial{\CW} + \initial{\CQ} \, , \qquad \beta := \frac{\initial{\CW}}{3} - \initial{\CQ} \, , \qquad \gamma := \frac{\initial{\CW}}{3} + \initial{\CQ} = \frac{1}{2} (\alpha - \beta) \, .
\end{equation}

\noindent
The Chaplygin fluid constraints imply $\alpha < 0$ and $\beta < 0$. The evolution of ${\CQ}_\CD$ and ${\CW}_\CD$ is entirely determined by the initial values $\initial{\CQ}$ and $\initial{\CW}$ of the domain or,~equivalently,~by $\alpha$ and $\beta$. The opposite sign of $\gamma$ gives the sign of $B_\CD$, equation (\ref{Binit}), and therefore defines the behaviour of the Chaplygin backreaction fluid.


\subsection{Evolution of the curvature deviation} \label{subs-limitW}

The rate of change $\partial {\CW}_\CD / \partial a_\CD$ shows that the curvature deviation grows with $a_\CD$. For large values of the scale factor, $a^{6}_\CD \gg |\gamma / \beta|$, we get

\begin{equation}
	{\CW}_\CD \sim -\frac{3}{2} \, (\alpha \beta)^{1/2} \, ,
\end{equation}

\noindent
which takes the form, using equation (\ref{Ainit}),

\begin{equation}
	{\CW}_\CD \sim - 24 \pi G \sqrt{A_\CD} \, = \, \CW_M \, ,
\end{equation}

\noindent
where $\CW_M$ has been introduced before. Thus,~for a given $A_\CD$,~the curvature deviation increases towards its attractor $\CW_M$, for which the departure from the Friedmannian curvature is maximum. For $\gamma = 0$ the curvature deviation is initially set to $\CW_M$ and then does not evolve. It is important to remark that there exists an infinite number of couples $(\initial{\CQ},\initial{\CW})$ or,~equivalently,~$(\alpha,\beta)$ which yield the same $A_\CD$ under the Chaplygin fluid constraints. For all of them $\CW_\CD$ will live on the same iso--$A_\CD$ curve and will tend towards the same attractor (see figure \ref{fig:iso-A}).


\subsection{Evolution of the backreaction} \label{subs-limitQ}

For large values of the scale factor, $a^{6}_\CD \gg |\gamma / \beta|$, we have

\begin{equation}
	{\CQ}_\CD \sim \, \frac{1}{2} \, \big( \alpha \beta \big)^{1/2} \, ,
\end{equation}

\noindent
or, equivalently, using equation (\ref{Ainit}),

\begin{equation}
	{\CQ}_\CD \sim \, 8 \pi G \sqrt{A_\CD} \, = \, \CQ_{\CW_M} > 0 \, ,
\end{equation}

\noindent
where $\CQ_{\CW_M}$ has been introduced before. The backreaction tends at late times towards a domain--dependent cosmological constant whose value is given by the initial conditions $\alpha$ and $\beta$ on the domain. The same remark as in subsection \ref{subs-limitW} can be made for ${\CQ}_\CD$. One needs to distinguish the following cases (see figure \ref{fig:iso-A}).

\begin{enumerate}
	\item For $\gamma > 0$ the backreaction is always positive and hence only acts as dark energy over the domain. In this situation, since $\partial {\CQ}_\CD / \partial a_\CD$ is negative, ${\CQ}_\CD$ behaves as dark energy whose intensity decreases until reaching the attractor $\CQ_{\CW_M}$. We recall in this case the existence of a minimal scale factor $a^{\mathrm{min}}_\CD = (- 2 \gamma / \beta)^{1/6}$. $\alpha < 0$ implies $a^{\mathrm{min}}_\CD < 1$.
	\item For $\gamma = 0$ the system is initially set on the attractor; thus it does not evolve. The backreaction always acts as a cosmological constant.
	\item for $\gamma < 0$ we have a positive $\partial {\CQ}_\CD / \partial a_\CD$. Two subcases arise according to the initial value $\initial{\CQ}$:
		\begin{enumerate}
			\item for $0 \leq \initial{\CQ} < - \initial{\CW} / 3$ the backreaction acts as dark energy whose intensity increases until reaching the attractor $\CQ_{\CW_M}$;
			\item for $\initial{\CW} / 3 < \initial{\CQ} < 0$ the backreaction changes its sign during its evolution. It first behaves as dark matter whose intensity decreases, and then as dark energy whose intensity increases until reaching the attractor. The moment of the transition dark matter--dark energy depends on the initial values of the domain since $a^{\mathrm{tr}}_\CD = (\gamma / \beta)^{1/6}$.
		\end{enumerate}
\end{enumerate}

\noindent
We conclude by emphasizing some of the characteristics of the model. First, the CG relates the backreaction terms of our inhomogeneous universe model and therefore furnishes a particular manifestation of its inhomogeneous structure. Our model does not suppose the existence of any other fundamental field, contrary to the approach of a FLRW universe filled with dust and CG. Second, the role and the evolution of the backreaction depend on the domain considered and on its initial values. As a consequence this model could be seen as an effective multi--scale model. For instance ${\CQ}_\CD$ might act as dark matter on small scales (situation (iii)-(b)) with a dark energy transition occuring at very late times, and as different kinds of dark energy on larger scales (situations (i), (ii) and (iii)-(a)).


\subsection{Another approach: the backreaction fluid as a scalar field} \label{bcksf}

The backreaction fluid might also be described by a minimally coupled real scalar field $\phi_\CD$,~called the morphon field \cite{morphon},~evolving in an effective potential $U_\CD(\phi_\CD)$,~as follows:

\begin{equation} \label{morphfield}
	\varrho_{\phi}^{\CD} : = \frac{\epsilon}{2} \dot{\phi}^2_{\CD} + U_{\CD}  \, , \qquad p_{\phi}^{\CD} : = \frac{\epsilon}{2} \dot{\phi}^2_{\CD} - U_{\CD} \, ,
\end{equation}

\noindent
where $\epsilon = + 1$ for a standard scalar field (with a positive kinetic energy) and $\epsilon = - 1$ for a phantom scalar field (with a negative kinetic energy). The last expressions together with relations (\ref{back-fluid}) give

\begin{equation} \label{pot-field}
	\epsilon \dot{\phi}^2_{\CD} = - \frac{1}{8 \pi G} \, ({\CQ}_\CD + \frac{{\CW}_\CD}{3}) \, , \qquad U_{\CD} = - \frac{1}{24 \pi G}  {\CW}_\CD \, .
\end{equation}

\noindent
The system evolves towards the maximal value of the curvature deviation $\CW_M$, as seen in subsection \ref{subs-limitW}, which corresponds to the minimal value of the potential. With correspondence (\ref{pot-field}) the integrability condition (\ref{integrability}) implies that $\phi_{\CD}$, for $\dot{\phi}_{\CD} \ne 0$, obeys the scale--dependent Klein--Gordon equation

\begin{equation} \label{klein-gordon}
	\ddot{\phi}_{\CD} + 3 \frac{\dot{a}_\CD}{a_\CD} \, \dot{\phi}_{\CD} + \epsilon \frac{\partial}{\partial \phi_{\CD}} U_{\CD} = 0 \, .
\end{equation}

\noindent
The above correspondence allows us to interpret the kinematical backreaction effects in terms of the properties of scalar field cosmologies, notably quintessence or phantom--quintessence scenarii that are here routed back to models of inhomogeneities. The morphon field may also be characterized by the domain--dependent equation of state $p^{\CD}_\phi = w^{\CD}_\phi \varrho^{\CD}_\phi \,$,~which assumes in our model the form

\begin{equation} \label{coses}
	w^{\CD}_{\phi}  \, = \,  \frac{-1}{1 \, + \, 2 \, {\displaystyle \frac{\gamma}{\beta}} \, a_{\CD}^{-6}} \, .
\end{equation}

\noindent
For $\gamma < 0 \,$,~the morphon field acts as dark matter and `standard' dark energy as mentioned in subsection \ref{posB}. In this situation we have $-1 < w^{\CD}_{\phi} < 0$ and $w^{\CD}_{\phi} \rightarrow -1^{+}$ at late times. For $\gamma > 0 \,$,~it behaves as phantom dark energy as mentioned in subsection \ref{negB}. In this case we have,~since $a_{\CD} > a^{\mathrm{min}}_\CD$, $w^{\CD}_{\phi} < -1$ and $w^{\CD}_{\phi} \rightarrow -1^{-}$ at late times. We stress again that the phantom character is an effective property in our description;~no fundamental phantom field is assumed to exist \cite{morphon}. Finally,~for $\gamma = 0 \,$,~the morphon field mimics a scale--dependent cosmological constant and we have $w^{\CD}_{\phi} = -1$.

Using equations (\ref{evol-r}) and (\ref{evol-q}), relations (\ref{morphfield}) become

\vspace{-5mm}
\begin{eqnarray}
	\epsilon \dot{\phi}^2_{\CD} & = & \frac{1}{8 \pi G} \, \frac{\alpha \gamma \, a^{-6}_{\CD}}{\left(\alpha \beta + 2 \, \alpha \gamma \, a^{-6}_{\CD} \, \right)^{1/2}} \, , \\
	U_{\CD} & = & \frac{1}{16 \pi G} \, \frac{\alpha \beta + \alpha \gamma \, a^{-6}_{\CD}}{\left(\alpha \beta + 2 \, \alpha \gamma \, a^{-6}_{\CD} \, \right)^{1/2}} \, .
\end{eqnarray}

\noindent
The scalar field dynamics can be reconstructed by evaluating the following integral, e.g.\ to find $U_{\CD}(\phi_\CD)$:

\begin{equation} \label{scalar_field}
{\fl
	\epsilon (\phi^{'}_{\CD})^2 = \frac{3}{4 \pi G} \, \frac{\alpha \gamma \, a_{\CD}^{-8}}{\Big[ 16 \pi G {\displaystyle \frac{M}{V_{\initial\CD}}} a_{\CD}^{-3} + \left(\alpha \beta + 2 \, \alpha \gamma \, a^{-6}_{\CD} \right)^{1/2} - 6 k_{\CD_{\it i}} a_{\CD}^{-2} \Big] \Big(\alpha \beta + 2 \, \alpha \gamma \, a^{-6}_{\CD} \Big)^{1/2}}  \, ,
}
\end{equation}

\noindent
where the prime denotes the derivation w.r.t.\ the volume scale factor. This relation does not seem analytically integrable in the general case. In the vacuum with a zero--Friedmaniann curvature ($k_{\CD_{\it i}} = 0$), expression (\ref{scalar_field}) becomes

\begin{equation} \label{field_vacuum}
	\epsilon (\phi^{'}_{\CD})^2 \, = \, \frac{3}{4 \pi G} \, \frac{\alpha \gamma \, a_{\CD}^{-8}}{\alpha \beta + 2 \, \alpha \gamma \, a^{-6}_{\CD} }  \, .
\end{equation}

\noindent
From this last relation, we get the following expression for a standard scalar field ($\epsilon = +1$ and $\gamma < 0$):

\begin{equation}
	\phi_{\CD}(a_{\CD}) \, = \, \mp \frac{1}{\sqrt{24 \pi G}} \, \mathrm{arccosh} \, \sqrt{\frac{A + B a^{-6}_{\CD}}{A}} + \phi_0 \, ,
\end{equation}

\noindent
where $\phi_0$ is an integration constant. The potential is then written \cite{kam:alter}

\begin{equation} \label{stand_vaccum}
{\fl
	U_{\CD}(\phi_{\CD}) \, = \, \frac{1}{32 \pi G} \, \sqrt{\alpha \beta} \, \left(\cosh{\sqrt{24 \pi G} (\phi_{\CD} - \phi_0)} + \frac{1}{\cosh{\sqrt{24 \pi G} (\phi_{\CD} - \phi_0)}} \right) \, .
}
\end{equation}

\noindent
For a phantom scalar field ($\epsilon = -1$ and $\gamma > 0$) we get from the relation (\ref{field_vacuum})

\begin{equation}
	\phi_{\CD}(a_{\CD}) \, = \, \mp \frac{1}{\sqrt{24 \pi G}} \, \arccos{\sqrt{\frac{A + B a^{-6}_{\CD}}{A}}} + \phi_0 \, ,
\end{equation}

\noindent
and the potential reads

\begin{equation}
{\fl
	U_{\CD}(\phi_{\CD}) \, = \, \frac{1}{32 \pi G} \, \sqrt{\alpha \beta} \, \left(\cos{\sqrt{24 \pi G} (\phi_{\CD} - \phi_0)} + \frac{1}{\cos{\sqrt{24 \pi G} (\phi_{\CD} - \phi_0)}} \right) \, .
}
\end{equation}

\noindent
Note that this last expression can be obtained directly from relation (\ref{stand_vaccum}) by the simple reparametrization $\phi_{\CD} - \phi_0 \rightarrow i (\phi_{\CD} - \phi_0)$. This is obvious in view of the form of the kinetic term of the morphon field: for a standard scalar field it is written $\dot{\phi}^2_{\CD}$ and for a phantom field $-\dot{\phi}^2_{\CD}$.

In the following section, we reformulate the different results obtained in this section in terms of effective cosmological parameters and we study the dynamics of the model.


\section{Effective cosmological parameters} \label{sec-cosparm}


\subsection{Constraints and evolution equations for the cosmological parameters}

Expressed through the domain--dependent cosmological parameters

\vspace{-5mm}
\begin{eqnarray}
	\Omega_m^{\CD} : = \frac{8 \pi G}{3 H_{\CD}^2} \langle\varrho\rangle_{\cal D} \, , & \qquad & \Omega_k^{\CD} : = - \frac{k_{\CD_{\it i}}}{H_{\CD}^2 \, a^2_\CD} \, , \\
	\Omega_{\CW}^{\CD} := - \frac{{\CW}_{\CD}}{6 H_{\CD}^2 } \, , & \qquad & \Omega_{\CQ}^{\CD} := - \frac{{\CQ}_{\CD}}{6 H_{\CD}^2 } \, ,
\end{eqnarray}

\noindent
where $H_{\CD} = \dot{a}_\CD/a_\CD$ is the volume Hubble functional, the averaged Hamiltonian constraint (\ref{eff-fried-1}) assumes the form \cite{buchert:review}

\begin{equation} \label{hamiltonomega}
	\Omega_m^{\CD} \, + \, \Omega_k^{\CD} \, + \, \Omega_{\CW}^{\CD} \, + \, \Omega_{\CQ}^{\CD} \, = \, 1 \, .
\end{equation}

\noindent
The Chaplygin fluid constraints (\ref{ChapconstI}) may be expressed as 

\begin{equation} \label{ChapconstII}
	\Omega_{\CW}^{\CD} \, + \, \Omega_{\CQ}^{\CD} \, > \, 0 \, , \qquad \frac{\Omega_{\CW}^{\CD}}{3} \, - \, \Omega_{\CQ}^{\CD} \, > \, 0 \, .
\end{equation}

\noindent
Equations (\ref{evol-r}) and (\ref{evol-q}) furnish the evolution laws for $\, \Omega_{\CW}^{\CD} \,$ and $\, \Omega_{\CQ}^{\CD} \,$:

\vspace{-5mm}
\begin{eqnarray}
	\Omega_{\CW}^{\CD} & \, = \, & \; \frac{3}{2} \; \frac{H_i^2}{H_{\CD}^2} \; \frac{\Omalpha \Ombeta \, + \, \Omalpha \Omgamma \, a_\CD^{-6} }{\left( \Omalpha \Ombeta + 2 \, \Omalpha \Omgamma \, a_\CD^{-6} \, \right)^{1/2}} \, , \label{evol-omr} \\
	\Omega_{\CQ}^{\CD} & \, = \, & - \frac{1}{2} \; \frac{H_i^2}{H_{\CD}^2} \; \frac{\Omalpha \Ombeta \, - \, \Omalpha \Omgamma \, a_\CD^{-6}}{\left( \Omalpha \Ombeta + 2 \, \Omalpha \Omgamma \, a_\CD^{-6} \, \right)^{1/2}} \, , \label{evol-omq}
\end{eqnarray}

\noindent
where the new terms are defined by

\begin{equation}
	\Omalpha := - \frac{\alpha}{6 H_i^2} \, , \qquad \Ombeta := - \frac{\beta}{6 H_i^2} \, , \qquad \Omgamma := - \frac{\gamma}{6 H_i^2} = \frac{1}{2} \, (\Omalpha - \Ombeta) \, .
\end{equation}

\noindent
The sign of $\Omgamma$ determines the behaviour of $\Omega_{\CW}^{\CD}$ and $\Omega_{\CQ}^{\CD}$. One may reformulate constraints (\ref{ChapconstII}) in terms of the initial conditions,

\begin{equation}
	0 \, < \, \Omalpha \, , \qquad 0 \, < \, \Ombeta \, .
\end{equation}


\subsection{Evolution of the Hubble functional}

Equation (\ref{hamiltonomega}) together with equations (\ref{evol-omr}) and (\ref{evol-omq}) provide the evolution equation for the Hubble functional (see appendix A for the study of $H_{\CD}$ in a particular case):

\begin{equation} \label{hubfunc}
	\frac{H_{\CD}^2}{H_i^2} \, = \, \Omega_k^i \, a_\CD^{-2} \, + \, \Omega_m^i \, a_\CD^{-3} \, + \, \left(\Omalpha \Ombeta + 2 \, \Omalpha \Omgamma \, a_\CD^{-6} \, \right)^{1/2} \, .
\end{equation}

\noindent
$H_{\CD}$ tends, at late times, towards

\begin{equation}
	\frac{H_{\CD}^2}{H_i^2} \, \sim \, \left(\Omalpha \Ombeta \right)^{1/2} \, .
\end{equation}


\subsection{Evolution of the cosmological density parameters} \label{evol-param}

One finds, for large values of the scale factor,

\begin{equation}
	\Omega_{\CW}^{\CD} \, \sim \, \frac{3}{2} \, , \qquad \Omega_{\CQ}^{\CD} \, \sim \, - \frac{1}{2} \, .
\end{equation}

\noindent
It is interesting to note that the cosmological parameters tend towards a value independent of the initial conditions (see appendix A for the study of $\Omega_{\CW}^{\CD}$ and $\Omega_{\CQ}^{\CD}$  in a particular case). We may also introduce the cosmological parameter $\Omega_X^{\CD}$, the so--called X--matter, defined as

\begin{equation}
	\Omega_X^{\CD} \, := \, \Omega_{\CW}^{\CD} \, + \, \Omega_{\CQ}^{\CD} \, = \, \frac{H_i^2}{H_{\CD}^2} \, \left( \Omalpha \Ombeta + 2 \, \Omalpha \Omgamma \, a_\CD^{-6} \, \right)^{1/2} \, .
\end{equation}

\noindent
We have noticed in section \ref{sec-inhomcosm} that the kinematical backreaction might explain the origin of dark matter and dark energy, and we have presented its behaviour in subsection \ref{subs-limitQ}. If one wants to compare our model to a scale--dependent Friedmannian cosmology in terms of cosmological parameters,~$\Omega_X^{\CD}$ has to be considered as the origin of the dark components, instead of $\Omega_{\CQ}^{\CD}$ alone. This simply means that the curvature deviation,~since it also participates in the departure from the Friedmannian framework,~may also act,~qualitatively,~as the dark components. In fact we know from previous work that the curvature deviation is actually quantitatively more important than the backreaction term itself \cite{buchert:review}.

\begin{figure}[ht]
	\begin{center}
	\vspace{1cm}
	\includegraphics[scale=1]{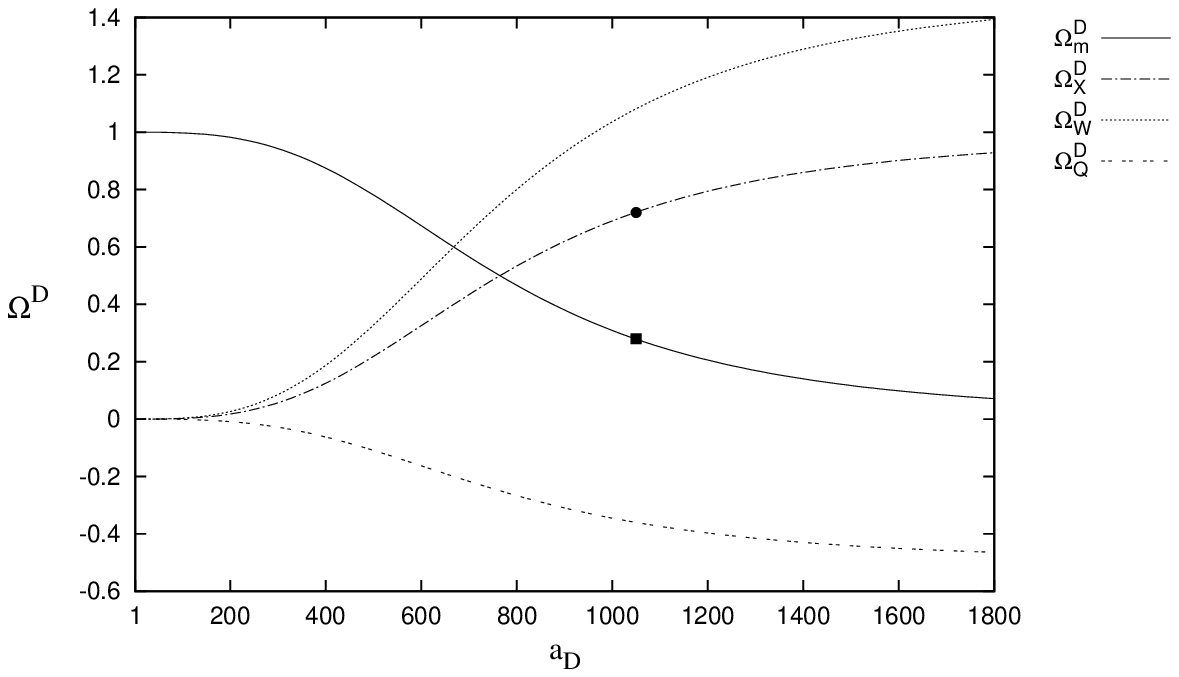}
	\includegraphics[scale=1]{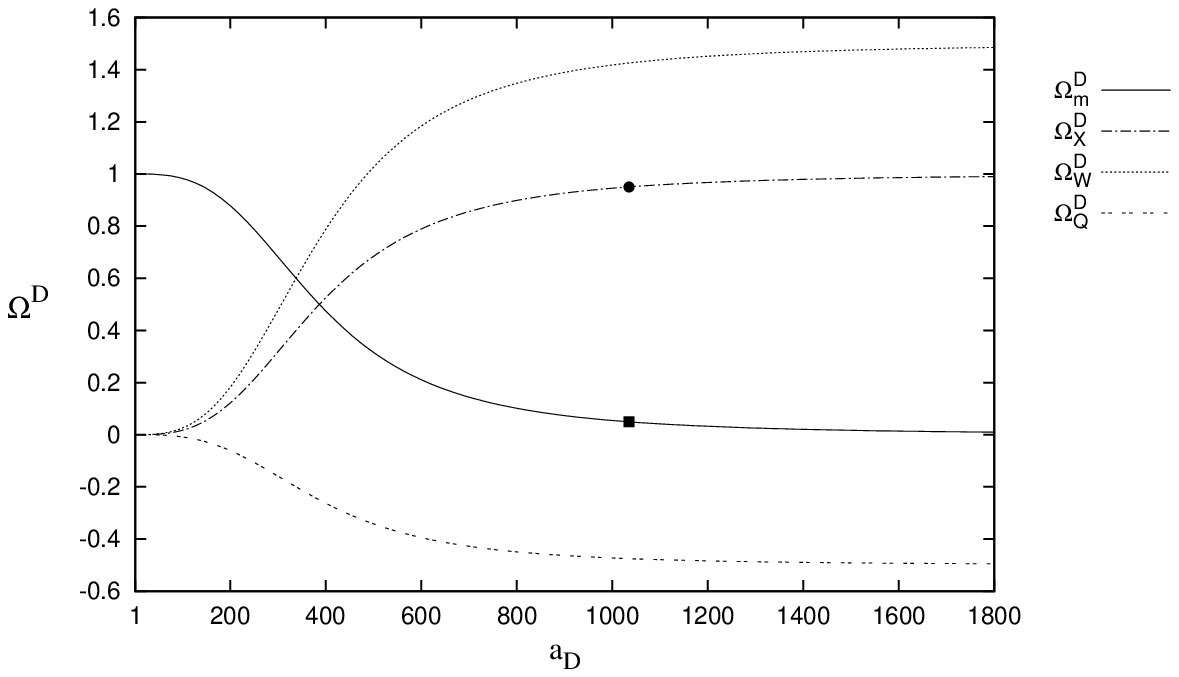}
	\caption{Evolution of the cosmological parameters $\Omega_m^{\CD}$, $\Omega_X^{\CD}$, $\Omega_{\CW}^{\CD}$ and $\Omega_{\CQ}^{\CD}$ w.\ r.\ t.\ the scale factor. We set $\Omega_k^{\CD} = 0$, $\Omega_m^i = 1-10^{-5}$ and $\Omega_X^i = \Omalpha = 10^{-5}$ where the initial moment is the CMB epoch. Upper figure: $\Ombeta = 5 \cdot 10^{-13}$; we have $\Omega_X^{\CD}(a_\CD \sim 1000) = \Omega_{DE}^0 \sim 0.72$ (dot) and $\Omega_m^{\CD}(a_\CD \sim 1000) = \Omega_b^0 + \Omega_{DM}^0 \sim 0.28$ (square). Lower figure: $\Ombeta = 3 \cdot 10^{-11}$; we have $\Omega_X^{\CD}(a_\CD \sim 1000) = \Omega_{DE}^0 +  \Omega_{DM}^0 \sim 0.95$ (dot) and $\Omega_m^{\CD}(a_\CD \sim 1000) = \Omega_b^0 \sim 0.05$ (square).}
	\label{param-cosmo}
	\end{center}
\end{figure}


We depict in figure \ref{param-cosmo} two situations to illustrate our model. The initial moment is chosen to be the epoch of the CMB, and we set $\Omega_k^{\CD} = 0$\footnote{This choice allows us to roughly compare our model to the concordance model. In this situation we have $\CW_{\CD} = \average{\CR}$.}, $\Omega_m^i = 1-10^{-5}$ and $\Omega_X^i = 10^{-5}$. In the upper figure, the X--matter only stands for dark energy; $\Omega_X^{\CD}$ corresponds to $\Omega_{DE}^F$ (the dark energy in the concordance model) and $\Omega_m^{\CD}$ to $\Omega_b + \Omega_{DM}^F$ (respectively the baryons and the dark matter in the concordance model). In the lower figure, the X--matter stands for dark energy and dark matter (since it can play both roles, it is regarded in this case as a result of different contributions on different scales); $\Omega_X^{\CD}$ corresponds to $\Omega_{DE}^F + \Omega_{DM}^F$  and $\Omega_m^{\CD}$ to $\Omega_b$. The two situations describe a universe initially close to a homogeneous--isotropic Friedmannian universe ($\Omega_X^i \sim 0$) with zero--curvature ($\Omega_k^{\CD} = 0$). This low deviation, however, becomes larger with the growth of the scale factor to reach in the first case $\Omega_X^{\CD}(a_\CD \sim 1000) = \Omega_X^0 = \Omega_{DE}^0 \sim 0.72$ (where the superscript denotes the `today'--value), and in the second case $\Omega_X^{\CD}(a_\CD \sim 1000) = \Omega_X^0 = \Omega_{DE}^0 + \Omega_{DM}^0 \sim 0.95$\footnote{If we choose the initial moment to be the CMB epoch, $a_{\CD} \sim 1000$ is chosen to roughly correspond to the `today'--value of the scale factor in a Friedmannian cosmology. Under the assumption that the metric of our universe does not significantly differ from a Euclidean metric, the scale factor,~calculated through this metric (equation (\ref{scale-factor})), evolves in our model approximately like its Friedmannian counterpart. Thus, we are entitled to assume $a_{\CD}^0 \sim 1000$. However, backreaction terms, encoding the inhomogeneities, involve first and second derivatives of the metric and can therefore not be neglected, even if the metrical amplitudes are considered to be near--Euclidean (see the discussion in \cite{estim} and numerical applications therein).}. The origin and magnitude of the dark energy (first situation) and of both dark energy and dark matter (second situation) are then entirely explained by the particular geometrical structure of an inhomogeneous space obeying the Chaplygin equation of state.

\begin{figure}[ht]
	\begin{center}
	\vspace{1cm}
	\includegraphics[scale=1]{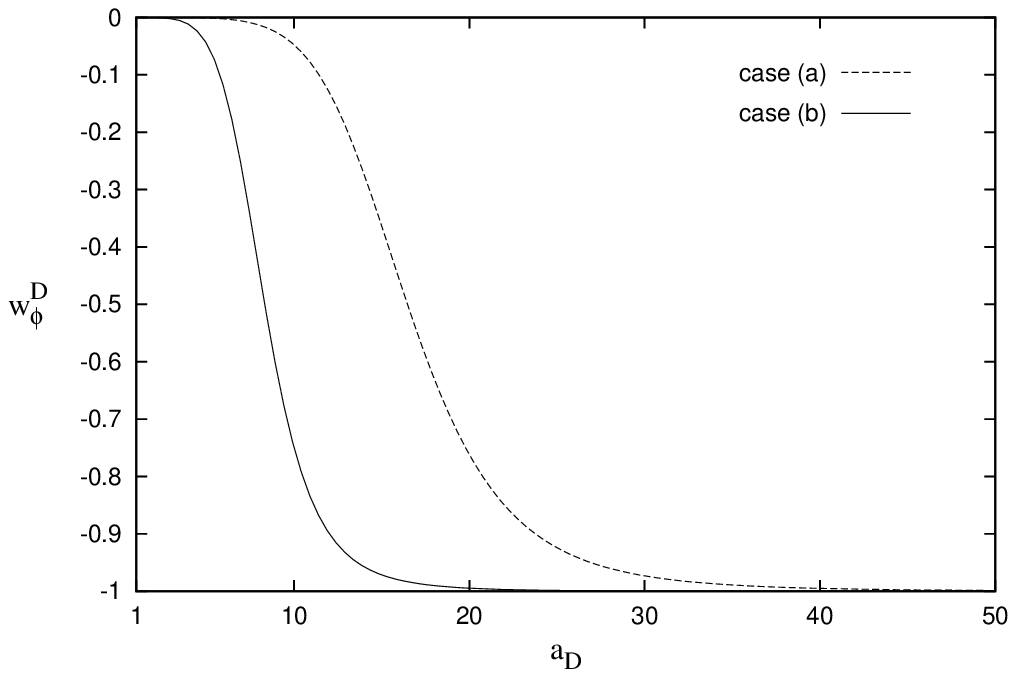}
	\caption{Evolution of $w^{\CD}_{\phi}$ w.\ r.\ t.\ the scale factor. $\Omega_k^{\CD} = 0$, $\Omega_m^i = 1-10^{-5}$ and $\Omega_X^i = \Omalpha = 10^{-5}$; the initial moment is the CMB epoch. Case (a): $\Ombeta = 5 \cdot 10^{-13}$ and $\Omega_X^{\CD} = \Omega_{DE}^F$. Case (b): $\Ombeta = 3 \cdot 10^{-11}$ and $\Omega_X^{\CD} = \Omega_{DE}^F + \Omega_{DM}^F$. In both situations $w^{\CD}_{\phi}$ quickly evolves towards $-1$, which corresponds to the cosmological constant--like behaviour of the backreaction fluid.}
	\label{w}
	\end{center}
\end{figure}


\begin{figure}[ht]
	\begin{center}
	\vspace{1cm}
	\includegraphics[scale=1]{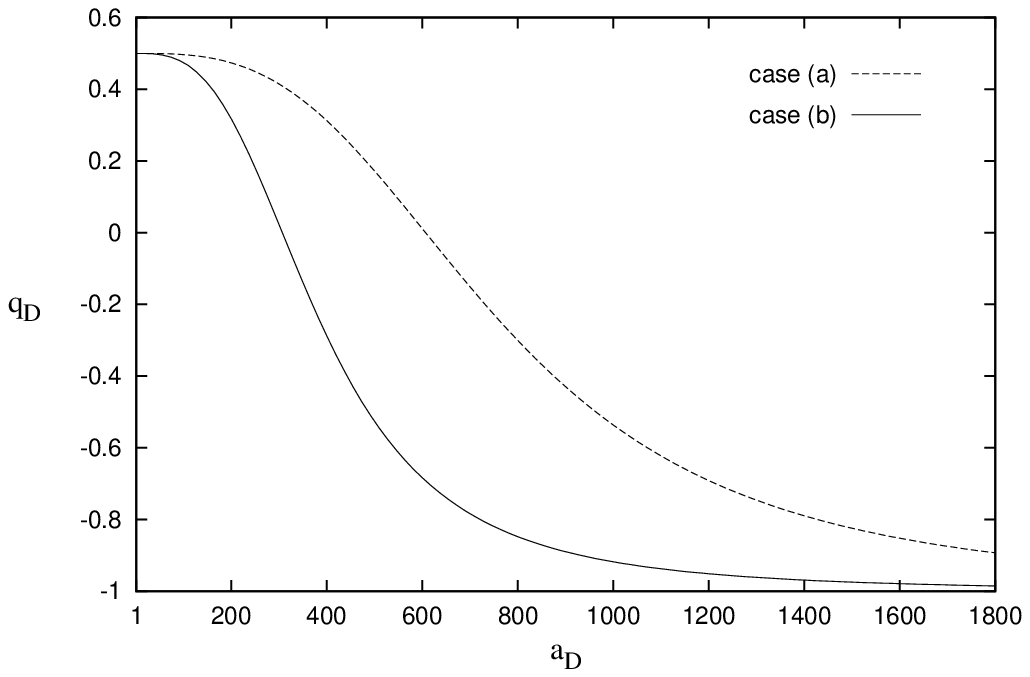}
	\caption{Evolution of the deceleration parameter w.\ r.\ t.\ the scale factor. $\Omega_k^{\CD} = 0$, $\Omega_m^i = 1-10^{-5}$ and $\Omega_X^i = \Omalpha = 10^{-5}$; the initial moment is the CMB epoch. Case (a): $\Ombeta = 5 \cdot 10^{-13}$ and $\Omega_X^{\CD} = \Omega_{DE}^F$. Case (b): $\Ombeta = 3 \cdot 10^{-11}$ and $\Omega_X^{\CD} = \Omega_{DE}^F + \Omega_{DM}^F$. The evolution of the deceleration parameter is only shaped by $\Omega_m^{\CD}$ and $\Omega_{\CQ}^{\CD}$ ($\Omega_X^{\CD}$ does not contribute to it, see equation (\ref{eq:q})). In both situations the expansion of the domain is first decelerated ($q_{\CD} > 0$), then accelerated ($q_{\CD} < 0$).}
	\label{dec-param}
	\end{center}
\end{figure}

\subsection{Another approach: the morphon equation of state}

Using the effective cosmological parameters, we rewrite equation (\ref{coses}) as

\begin{equation}
	w^{\CD}_{\phi} \, = \, \frac{- 1}{1 \, + \, 2 \, {\displaystyle \frac{\Omgamma}{\Ombeta}} \, a_{\CD}^{-6}} \, .
\end{equation}

\noindent
The behaviour of $w^{\CD}_{\phi}$ is known in terms of cosmological parameters simply by replacing $\gamma$ with $- \, \Omgamma$ through subsection \ref{bcksf} (see also figure \ref{w}).

\subsection{Dynamics of the fluid volume}

Depending on the initial conditions, the domain $\CD$ might undergo a decelerated or an accelerated expansion at different periods of its evolution. To learn how it behaves we throw an eye on the volume deceleration parameter

\begin{equation} \label{eq:q}
	q_{\CD} \, = \, - \frac{1}{H_{\CD}^2} \, \frac{\ddot{a}_{\CD}}{a_{\CD}} \, = \, \frac{1}{2} \, \Omega_m^{\CD} \, + \, 2 \, \Omega_{\CQ}^{\CD} \, .
\end{equation}

\noindent
At late times, since $q_{\CD} \, \sim \, -1$, the fluid volume undergoes an accelerated expansion, we are indeed dealing with a cosmological constant in this situation. Three different cases occur for the dynamics of the domain (see figure \ref{dec-param}, and appendix A for a detailed analysis in a particular case):

\begin{enumerate}
	\item its expansion is first decelerated then accelerated;
	\item its expansion is always accelerated;
	\item its expansion is accelerated, then decelerated and again accelerated.
\end{enumerate}

The latter case is an interesting situation: the kinematical backreaction is responsible for two accelerated phases. Our model only concerns the matter--dominated universe; however this situation allows to imagine that the primordial inflation and the one occuring today might be driven by the same effect studied in a more general model.


\section{Concluding Remarks} \label{sec:concl}

We have built in this paper a model in which the backreaction fluid acts as a domain--dependent effective CG, focusing for simplicity on the case where its effective energy density is positive. The behaviour of the kinematical backreaction only depends on the initial conditions of the domain and may correspond to dark matter or different types of dark energy according to the scale and the time evolution. Our model kinematically resembles a Friedmannian cosmology with two fluids (dust and CG), but conceptual implications differ. First, the origin of dark energy (or, in the extreme case, of both dark components) is related to the non--trivial geometrical structure of an inhomogeneous space; we do not assume the existence of other fundamental constituents or fields. Second, the geometrical structure that complies with the Chaplygin equation of state can in principle be verified by concrete inhomogeneous models and also observationally; the model is no longer phenomenological in the sense that no free parameters remain. Free parameters in the standard approach (being homogeneous) are here traced back to the initial data for the inhomogeneities and are therefore, in this sense, not arbitrary. Any fitting of observational data of our model will lead to unambiguous initial data that can be constraint by structure formation. Third, one has to be careful when cosmological observables are derived. Indeed, angular diameter and luminosity distances, for instance, depend on metrical properties. These latter are affected and related, in our description, to the averaged scalar curvature, which evolves differently compared with its Friedmannian counterpart. It is therefore necessary to reinterprete observational data, using e.g.\ the lines of the analysis performed for the exact scaling solutions in \cite{morphon:obs}. In light of this remark it is premature to exclude \cite{amen,bento2,bento3,cun} the standard Chaplygin equation of state as providing a good match with observational data. The construction of an effective metric for a cosmology with a Chaplygin backreaction fluid and a comparison with observations are the subjects of a future work. 

\vspace{15pt}

\noindent
{Acknowledgements:}

\noindent
{\footnotesize
We wish to thank Charly Nayet and Brice Riba for useful discussions. Also, we wish to thank Ugo Moschella, Alexander Kamenshchik, Roberto Sussman and Gary Tupper for their useful comments on the manuscript and interesting suggestions. 
This work was supported by `F\'ed\'eration de Physique Andr\'e Marie Amp\`ere'. XR acknowledges support by \'Ecole Doctorale, Lyon. 
}


\renewcommand{\theequation}{A.\arabic{equation}}
\setcounter{equation}{0}

\section*{Appendix A. Study of the model in the case $\Omega_k^{\CD} = 0$} \label{appA}

Setting $\Omega_k^{\CD}$ to 0 allows to compare our model to the concordance model (we also suppose here that $\Omega_m^{\CD} \neq 0$ for any domain). In this situation equation (\ref{hamiltonomega}) becomes

\begin{equation}
	\Omega_m^{\CD} \, + \, \Omega_{\CW}^{\CD} \, + \, \Omega_{\CQ}^{\CD} \, = \, 1 \quad \Rightarrow \quad \Omega_m^i \, = \, 1 \, - \, \Omalpha \, .
\end{equation}

\noindent
Since $\Omega_m^{\CD}$ and $\Omega_{\CW}^{\CD} \, + \, \Omega_{\CQ}^{\CD}$ are both positive under the Chaplygin fluid constraints (see equation (\ref{ChapconstII})), one also needs

\begin{equation}
	\Omega_m^{\CD} \, < \, 1 \, , \qquad \Omega_{\CW}^{\CD} \, + \, \Omega_{\CQ}^{\CD} \, < \, 1 \, .
\end{equation}

\subsection*{A.1. Evolution of the Hubble functional}

Equation (\ref{hubfunc}) becomes

\begin{equation}
	H_{\CD}^2 \, = \, H_i^2 \left[ (1 - \Omalpha) \, a_\CD^{-3} \, + \, \left(\Omalpha \Ombeta + 2 \, \Omalpha \Omgamma \, a_\CD^{-6} \, \right)^{1/2} \right] \, .
\end{equation}

\noindent
For $\Omgamma \geq 0$ $H_{\CD}^2$ always decreases, and for $\Omgamma < 0$ it increases in the interval $(a_\CD^{\mathrm{min}},a_1)$ and decreases in $(a_1,+\infty)$ with

\begin{equation}
	(a_\CD^{\mathrm{min}})^{6} \, = \, -2 \, \frac{\Omgamma}{\Ombeta} \, , \qquad (a_1 )^6 \, = \, -2 \, \frac{\Omgamma}{\Ombeta} \, + \, 4 \, \frac{\Omalpha \, \Omgamma^2}{\Ombeta \, (1 - \Omalpha)^2} \, .
\end{equation}

\subsection*{A.2. Evolution of $\Omega_{\CW}^{\CD}$ and $\Omega_{\CQ}^{\CD}$}

The derivatives of equations (\ref{evol-omr}) and (\ref{evol-omq}) show that, for $\Omgamma \geq 0$, $\Omega_{\CW}^{\CD}$ increases and $\Omega_{\CQ}^{\CD}$ decreases. For $\Omgamma < 0$,  $\Omega_{\CW}^{\CD}$ decreases in $(a_\CD^{\mathrm{min}},a_2)$ and increases in $(a_2,+\infty)$, and $\Omega_{\CQ}^{\CD}$ increases in $(a_\CD^{\mathrm{min}},a_3)$ and decreases in $(a_3,+\infty)$, with

\vspace{-4mm}
\begin{eqnarray}
	\fl (a_2)^6 & \, = \, & \frac{- \Omgamma \left[3 (1 - \Omalpha)^2 - 2 \Omalpha \Omgamma \right] + 2 \sqrt{\Omalpha \Omega_{\gamma}^3 \left[\Omalpha \Omgamma - (1 - \Omalpha)^2 \right]}}{(1 - \Omalpha)^2 \, \Ombeta} \, , \\
	\fl (a_3)^6 & \, = \, & \frac{- \Omgamma \left[5 (1 \! - \! \Omalpha)^2 \! - \! 18 \Omalpha \Omgamma \right] \! + \! 6 \sqrt{3 \Omalpha \Omega_{\gamma}^3 \left[3 \Omalpha \Omgamma \! - \! (1 \! - \! \Omalpha)^2 \right]}}{(1 - \Omalpha)^2 \, \Ombeta} \, .
\end{eqnarray}

\subsection*{A.3. Evolution of the deceleration parameter}

We define the quantities

\vspace{-5mm}
\begin{eqnarray}
	\fl (a_4)^6 & \, := \, & - \frac{5}{2} \, \frac{\Omalpha - \Ombeta}{\Ombeta} \, , \\
	\fl (a_5)^6 & \, := \, & \frac{8 \, \Omalpha \Omgamma \, + \, (1 - \Omalpha)^2 \, - \, (1 - \Omalpha) \, \sqrt{(1 - \Omalpha)^2 \, + 48 \, \Omalpha \Omgamma}}{8 \, \Omalpha \Ombeta} \, , \\
	\fl (a_6)^6 & \, := \, & \frac{8 \, \Omalpha \Omgamma \, + \, (1 - \Omalpha)^2 \, + \, (1 - \Omalpha) \, \sqrt{(1 - \Omalpha)^2 \, + 48 \, \Omalpha \Omgamma}}{8 \, \Omalpha \Ombeta} \, , \\
	\fl \Omega_{{\alpha}_1} & \, := \, & \frac{12 \, \Ombeta \, + \, 1 \, - \, 2 \, \sqrt{6 \, (3 \, \Ombeta - 1) \, (2 \, \Ombeta + 1)}}{25} \, , \\
	\fl \Omega_{{\alpha}_2} & \, := \, & \frac{12 \, \Ombeta \, + \, 1 \, + \, 2 \, \sqrt{6 \, (3 \, \Ombeta - 1) \, (2 \, \Ombeta + 1)}}{25} \, , \\
	\fl \Omega_{{\alpha}_3} & \, := \, & \frac{1}{5} \, .
\end{eqnarray}

\noindent
and the situations

\begin{enumerate}[(a)]
	\item $\forall \, a_\CD \in (a_\CD^{\mathrm{min}}, + \infty) \;\; q_\CD < 0 \, $;
	\item $\forall \, a_\CD \in (a_\CD^{\mathrm{min}}, + \infty) \backslash \{a_4\} \;\; q_\CD < 0$, and $q_\CD (a_4) = 0 \, $;
	\item $\forall \, a_\CD \in (a_\CD^{\mathrm{min}}, a_5) \cup (a_6, + \infty) \;\; q_\CD < 0 \,$, \, $q_\CD (a_5) = q_\CD (a_6) = 0$, and $\forall \, a_\CD \in (a_5, a_6) \;\; q_\CD > 0 \, $;
	\item $\forall \, a_\CD \in (0, a_6) \;\; q_\CD > 0 \,$, $\,q_\CD (a_6) = 0$, and $\forall \, a_\CD \in (a_6, + \infty) \;\; q_\CD < 0 \, $.
\end{enumerate}

Table 1 presents the exact evolution of the deceleration parameter in these situations.

\begin{table}[ht]
\centering
	\begin{tabular}{|r|c|c|}
	\hline
	& $\scriptstyle (0, \, \Ombeta)$ & $\scriptstyle [\Ombeta,\, 1)$ \\
	\hline
	$\scriptstyle 0 \, < \, \Ombeta \, < \, 1/3$ & \scriptsize (c) & \scriptsize (d) \\
	\hline
	\end{tabular}
	\\[10pt]
	\begin{tabular}{|r|c|c|c|c|}
	\hline
	& $\scriptstyle (0, \, \Omega_{{\alpha}_3})$ & $\scriptstyle \Omega_{{\alpha}_3}$ & $\scriptstyle (\Omega_{{\alpha}_3}, \, \Ombeta)$ & $\scriptstyle [\Ombeta, \, 1)$ \\
	\hline
	$\scriptstyle \Ombeta = 1/3$ & \scriptsize (c) & \scriptsize (b) & \scriptsize (c) & \scriptsize (d) \\
	\hline
	\end{tabular}
	\\[10pt]
	\begin{tabular}{|r|c|c|c|c|c|c|}
	\hline
	& $\scriptstyle (0, \, \Omega_{{\alpha}_1})$ & $\scriptstyle \Omega_{{\alpha}_1}$ & $\scriptstyle (\Omega_{{\alpha}_1}, \, \Omega_{{\alpha}_2})$ & $\scriptstyle \Omega_{{\alpha}_2}$ & $\scriptstyle (\Omega_{{\alpha}_2}, \, \Ombeta)$ & $\scriptstyle [\Ombeta, \, 1)$ \\
	\hline
	$\scriptstyle 1/3 \, < \, \Ombeta \, < \, 1$ & \scriptsize (c) & \scriptsize (b) & \scriptsize (a) & \scriptsize (b) & \scriptsize (c) & \scriptsize (d) \\
	\hline
	\end{tabular}
	\\[10pt]
	\begin{tabular}{|r|c|c|c|}
	\hline
	& $\scriptstyle (0, \, \Omega_{{\alpha}_1})$ & $\scriptstyle \Omega_{{\alpha}_1}$ & $\scriptstyle (\Omega_{{\alpha}_1}, \, 1)$ \\
	\hline
	$\scriptstyle 1 \leq \, \Ombeta $ & \scriptsize (c) & \scriptsize (b) & \scriptsize (a) \\
	\hline
	\end{tabular}
	\caption{Evolution  of the deceleration parameter for the different values of $\Omalpha$ and $\Ombeta$ under the Chaplygin fluid constraints with $\Omega_k^{\CD} = 0$. In situations (a) and (b) the expansion of the domain is always accelerated;~in situation (c) it is first accelerated, then decelerated and again accelerated; in situation (d) it is first decelerated, then accelerated.}
\end{table}


\end{document}